\documentclass{article}
\begin{document}
{\LARGE
\begin{center}
{\bf
Scalar tetraquarks with open bottom}
\end{center}
}

\large

\begin{center}
\vskip3ex
S.M. Gerasyuta $ ^{*}$ and V.I. Kochkin

\vskip2ex
Department of Physics, LTA, 194021, St. Petersburg, Russia
\end{center}

\vskip2ex

\noindent
$ ^{*}$ E-mail: gerasyuta@SG6488.spb.edu

\vskip4ex
\begin{center}
{\bf Abstract}
\end{center}
\vskip4ex
{\large
The relativistic four-quark equations in the framework of the dispersion
relation technique are derived. The four-quark amplitudes of the scalar
tetraquarks with open bottom, including $u$, $d$, $s$ and bottom
quarks, are constructed. The poles of these amplitudes determine the masses
and widths of scalar tetraquarks.


\vskip2ex

\vskip2ex
{\bf I. Introduction.}
\vskip2ex
The observation of the $X(3872)$ [1 -- 4], the first of the $XYZ$ particles
to be seen, brought forward the hope that a multiquark state has been
received. Belle Collaboration discovered the $X(3940)$ in the reaction
$e^+ e^- \to J/ \psi +X$ [5]. The fact that the newly found states do not
fit quark model calculations [6] has triggered the interest in the
charmonium-like and the charmed states. Maiani et al. advocate a tetraquark
explanation for the $X(3872)$ [7, 8]. Ebert et al. [9] calculated that the
light scalar tetraquark lies above the open charm threshold and is broad,
while Maiani et al. obtained [10] that this state lies a few $MeV$ below
this threshold. In our paper [11] the tetraquark $X(3700)$ with the
spin-parity $J^{pc}=0^{++}$ decays to $\eta\eta_c$. The calculated width
of this state is equal to $\Gamma_{0^{++}}=140\, MeV$.

In the present paper the relativistic four-quark equations are found in
the framework of coupled-channel formalism. The dynamical mixing between the
meson-meson states and the four-quark states is considered [12 -- 14].
The masses and the widths of the scalar tetraquarks with open bottom
are calculated.

\vskip2ex
{\bf II. Four-Quark Amplitudes for the Tetraquarks with Open Bottom.}
\vskip2ex
We derive the relativistic four-quark equations in the framework of the
dispersion relations technique.

The correct equations for the amplitude are obtained by taking into
account all possible subamplitudes. It corresponds to the division of
complete system into subsystems with the smaller number of particles.
Then one should represent a four-particle amplitude as a sum of six
subamplitudes:

\begin{equation}
A=A_{12}+A_{13}+A_{14}+A_{23}+A_{24}+A_{34}\, . \end{equation}

This defines the division of the diagrams into groups according to the
certain pair interaction of particles. The total amplitude can be
represented graphically as a sum of diagrams.

We need to consider only one group of diagrams and the amplitude
corresponding, for example $A_{12}$.

The relativistic generalization of the Faddeev-Yakubovsky approach [15, 16]
for the tetraquark is obtained. We shall construct the four-quark amplitude
of $\bar b u \bar u u$ tetraquark with the spin-parity $J^{pc}=0^{++}$ in
which the quark amplitudes with quantum numbers of $0^{-+}$ and $1^{--}$
mesons are included. The set of diagrams associated with
the amplitude $A_{12}$ can further broken down into four groups
corresponding to subamplitudes: $A_1 (s,s_{12},s_{34})$,
$A_2 (s,s_{23},s_{14})$, $A_3 (s,s_{23},s_{123})$, $A_4 (s,s_{14},s_{124})$,
if we consider the tetraquark with the spin-parity $J^{pc}=0^{++}$
$(\bar b u \bar u u)$.

Here $s_{ik}$ is the two-particle subenergy squared, $s_{ijk}$ corresponds
to the energy squared of particles $i$, $j$, $k$ and $s$ is the system
total energy squared.

In order to represent the subamplitudes  $A_1 (s,s_{12},s_{34})$,
$A_2 (s,s_{23},s_{14})$, $A_3 (s,s_{23},s_{123})$ and
$A_4 (s,s_{14},s_{124})$ in the form of dispersion relations it is
necessary to define the amplitudes of quark-antiquark interaction
$a_n(s_{ik})$. The pair quarks amplitudes $q \bar q\rightarrow q \bar q$
are calculated in the framework of the dispersion $N/D$ method with the
input four-fermion interaction [17 -- 19] with quantum numbers of the
gluon [20]. The regularization of the dispersion integral for the
$D$-function is carried out with the cutoff parameter $\Lambda$.
The four-quark interaction is considered as an input [20]:

\begin{equation}
g_V \left(\bar q \lambda I_f \gamma_{\mu} q \right)^2 +
g^{(s)}_V \left(\bar q \lambda I_f \gamma_{\mu} q \right)
\left(\bar s \lambda \gamma_{\mu} s \right)+
g^{(ss)}_V \left(\bar s \lambda \gamma_{\mu} s \right)^2 \, . \end{equation}

\noindent
Here $I_f$ is the unity matrix in the flavor space $(u, d)$. $\lambda$ are
the color Gell-Mann matrices. Dimensional constants of the four-fermion
interaction $g_V$, $g^{(s)}_V$ and $g^{(ss)}_V$ are parameters of the
model. At $g_V =g^{(s)}_V =g^{(ss)}_V$ the flavor $SU(3)_f$ symmetry occurs.
The strange quark violates the flavor $SU(3)_f$ symmetry. In order to avoid
an additional violation parameters, we introduce the scale shift of the
dimensional parameters [20]:

\begin{equation}
g=\frac{m^2}{\pi^2}g_V =\frac{(m+m_s)^2}{4\pi^2}g_V^{(s)} =
\frac{m_s^2}{\pi^2}g_V^{(ss)} \, .\end{equation}

\begin{equation}
\Lambda=\frac{4\Lambda(ik)}
{(m_i+m_k)^2}. \end{equation}

\noindent
Here $m_i$ and $m_k$ are the quark masses in the intermediate state of
the quark loop. Dimensionless parameters $g$ and $\Lambda$ are supposed
to be constants which are independent of the quark interaction type. The
applicability of Eq. (2) is verified by the success of
De Rujula-Georgi-Glashow quark model [21], where only the short-range
part of Breit potential connected with the gluon exchange is
responsible for the mass splitting in hadron multiplets.

We use the results of our relativistic quark model [20] and write down
the pair quarks amplitude in the form:

\begin{equation}
a_n(s_{ik})=\frac{G^2_n(s_{ik})}
{1-B_n(s_{ik})} \, ,\end{equation}

\begin{equation}
B_n(s_{ik})=\int\limits_{(m_i+m_k)^2}^{\frac{(m_i+m_k)^2\Lambda}{4}}
\hskip2mm \frac{ds'_{ik}}{\pi}\frac{\rho_n(s'_{ik})G^2_n(s'_{ik})}
{s'_{ik}-s_{ik}} \, .\end{equation}

\noindent
Here $G_n(s_{ik})$ are the quark-antiquark vertex functions. The vertex
functions are determined by the contribution of the crossing channels.
The vertex functions satisfy the Fierz relations. All of these vertex
functions are generated from $g_V$, $g^{(s)}_V$ and $g^{(ss)}_V$.
$B_n(s_{ik})$, $\rho_n (s_{ik})$ are the Chew-Mandelstam functions with
cutoff $\Lambda$ and the phase spaces, respectively.

Here $n=1$ determines a $q \bar q$-pairs with $J^{pc}=0^{-+}$ in the $1_c$
color state, $n=2$ corresponds to a $q \bar q$-pairs with $J^{pc}=1^{--}$
in the $1_c$ color state, and $n=3$ defines the $q \bar q$-pairs
corresponding to tetraquarks with quantum numbers $J^{pc}=0^{++}$.

In the case in question, the interacting quarks do not produce a bound
state; therefore, the integration in Eqs. (7) -- (10) is carried out from
the threshold $(m_i+m_k)^2$ to the cutoff $\Lambda(ik)$.
The coupled integral equation systems (the tetraquark state with
$n=3$ and $J^{pc}=0^{++}$ $\bar b u \bar u u$) can be described as:

\begin{eqnarray}
A_1(s,s_{12},s_{34})&=&\frac{\lambda_1 B_2(s_{12})  B_2(s_{34})}
{[1- B_2(s_{12})][1- B_2(s_{34})]}+
2\hat J_2(s_{12},s_{34},2,2) A_3(s,s'_{23},s'_{123})\nonumber\\
&&\nonumber\\
&+&2\hat J_2(s_{12},s_{34},2,2) A_4(s,s'_{14},s'_{124}) \, ,\\
&&\nonumber\\
A_2(s,s_{23},s_{14})&=&\frac{\lambda_2 B_1(s_{23})  B_1(s_{14})}
{[1- B_1(s_{23})][1- B_1(s_{14})]}+
2\hat J_2(s_{23},s_{14},1,1) A_3(s,s'_{34},s'_{234})\nonumber\\
&&\nonumber\\
&+&2\hat J_2(s_{23},s_{14},1,1) A_4(s,s'_{12},s'_{123}) \, ,\\
&&\nonumber\\
A_3(s,s_{23},s_{123})&=&\frac{\lambda_3 B_3(s_{23})}{[1- B_3(s_{23})]}+
2\hat J_3(s_{23},3) A_1(s,s'_{12},s'_{34})
+\hat J_3(s_{23},3) A_2(s,s'_{12},s'_{34})\nonumber\\
&&\nonumber\\
&+&\hat J_1(s_{23},3) A_4(s,s'_{34},s'_{234})+
\hat J_1(s_{23},3) A_3(s,s'_{12},s'_{123}) \, ,\\
&&\nonumber\\
A_4(s,s_{14},s_{124})&=&\frac{\lambda_4 B_3(s_{14})}{[1- B_3(s_{14})]}+
2\hat J_3(s_{14},3) A_1(s,s'_{13},s'_{24})
+2\hat J_3(s_{14},3) A_2(s,s'_{13},s'_{24})\nonumber\\
&&\nonumber\\
&+&2\hat J_1(s_{14},3) A_3(s,s'_{14},s'_{134})
+2\hat J_1(s_{14},3) A_4(s,s'_{14},s'_{134}) \, ,
\end{eqnarray}

\noindent
where $\lambda_i$, $i=1, 2, 3, 4$ are the current constants. They do not
affect the mass spectrum of tetraquarks. We introduce the integral
operators:

\begin{eqnarray}
\hat J_1(s_{12},l)&=&\frac{G_l(s_{12})}
{[1- B_l(s_{12})]} \int\limits_{(m_1+m_2)^2}^{\frac{(m_1+m_2)^2\Lambda}{4}}
\frac{ds'_{12}}{\pi}\frac{G_l(s'_{12})\rho_l(s'_{12})}
{s'_{12}-s_{12}} \int\limits_{-1}^{+1} \frac{dz_1}{2} \, ,\\
&&\nonumber\\
\hat J_2(s_{12},s_{34},l,p)&=&\frac{G_l(s_{12})G_p(s_{34})}
{[1- B_l(s_{12})][1- B_p(s_{34})]}
\int\limits_{(m_1+m_2)^2}^{\frac{(m_1+m_2)^2\Lambda}{4}}
\frac{ds'_{12}}{\pi}\frac{G_l(s'_{12})\rho_l(s'_{12})}
{s'_{12}-s_{12}}\nonumber\\
&&\nonumber\\
&\times&\int\limits_{(m_3+m_4)^2}^{\frac{(m_3+m_4)^2\Lambda}{4}}
\frac{ds'_{34}}{\pi}\frac{G_p(s'_{34})\rho_p(s'_{34})}
{s'_{34}-s_{34}}
\int\limits_{-1}^{+1} \frac{dz_3}{2} \int\limits_{-1}^{+1} \frac{dz_4}{2}
 \, ,\\
&&\nonumber\\
\hat J_3(s_{12},l)&=&\frac{G_l(s_{12},\tilde \Lambda)}
{[1- B_l(s_{12},\tilde \Lambda)]} \, \, \frac{1}{4\pi}
\int\limits_{(m_1+m_2)^2}^{\frac{(m_1+m_2)^2\tilde \Lambda}{4}}
\frac{ds'_{12}}{\pi}\frac{G_l(s'_{12},\tilde \Lambda)
\rho_l(s'_{12})}
{s'_{12}-s_{12}}\nonumber\\
&&\nonumber\\
&\times&\int\limits_{-1}^{+1}\frac{dz_1}{2}
\int\limits_{-1}^{+1} dz \int\limits_{z_2^-}^{z_2^+} dz_2
\frac{1}{\sqrt{1-z^2-z_1^2-z_2^2+2zz_1z_2}} \, ,
\end{eqnarray}

\noindent
here $l$, $p$ are equal to $1-3$.

In Eqs. (11) and (13) $z_1$ is the cosine of the angle between the relative
momentum of particles 1 and 2 in the intermediate state and the momentum
of the particle 3 in the final state, taken in the c.m. of particles
1 and 2. In Eq. (13) $z$ is the cosine of the angle between the momenta
of particles 3 and 4 in the final state, taken in the c.m. of particles
1 and 2. $z_2$ is the cosine of the angle between the relative
momentum of particles 1 and 2 in the intermediate state and the momentum
of the particle 4 in the final state, is taken in the c.m. of particles
1 and 2. In Eq. (12): $z_3$ is the cosine of the angle between the relative
momentum of particles 1 and 2 in the intermediate state and the relative
momentum of particles 3 and 4 in the intermediate state, taken in the c.m.
of particles 1 and 2. $z_4$ is the cosine of the angle between the relative
momentum of particles 3 and 4 in the intermediate state and that of the
momentum of the particle 1 in the intermediate state, taken in the c.m.
of particles 3 and 4.

We can pass from the integration over the cosines of the angles to the
integration over the subenergies [22].

Let us extract two-particle singularities in the amplitudes
$A_1(s,s_{12},s_{34})$, $A_2 (s,s_{23},s_{14})$, $A_3 (s,s_{23},s_{123})$
and $A_4 (s,s_{14},s_{124})$:

\begin{equation}
A_1(s,s_{ik},s_{lm})=\frac{\alpha_1(s,s_{ik},s_{lm})B_2(s_{ik})B_2(s_{lm})}
{[1-B_2(s_{ik})][1-B_2(s_{lm})]} \, ,\end{equation}

\begin{equation}
A_2(s,s_{ik},s_{lm})=\frac{\alpha_2(s,s_{ik},s_{lm})B_1(s_{ik})B_1(s_{lm})}
{[1-B_1(s_{ik})][1-B_1(s_{lm})]} \, ,\end{equation}

\begin{equation}
A_j(s,s_{ik},s_{ikl})=\frac{\alpha_j(s,s_{ik},s_{ikl})B_3(s_{ik})}
{1-B_3(s_{ik})} \, , \quad\quad j=3-4 \, .\end{equation}

We do not extract three-particles singularities, because they are weaker
than two-particle singularities.

We used the classification of singularities, which was proposed in
paper [23]. The construction of the approximate solution of Eqs.
(7) -- (10) is based on the extraction of the leading singularities
of the amplitudes. The main singularities in $s_{ik}\approx (m_i+m_k)^2$
are from pair rescattering of the particles $i$ and $k$. First of all there
are threshold square-root singularities. Also possible are pole
singularities which correspond to the bound states. The amplitudes
apart from two-particle singularities have triangular singularities and the
singularities defining the interactions of four particles. Such
classification allows us to search the corresponding solution of Eqs.
(7) -- (10) by taking into account some definite number of leading
singularities and neglecting all the weaker ones. We consider the
approximation which defines two-particle, triangle and four-particle
singularities. The functions $\alpha_1(s,s_{12},s_{34})$,
$\alpha_2(s,s_{23},s_{14})$, $\alpha_3(s,s_{23},s_{123})$ and
$\alpha_4(s,s_{14},s_{124})$ are the smooth functions of $s_{ik}$,
$s_{ikl}$, $s$ as compared with the singular part of the amplitude, hence
they can be expanded in a series in the singulary point and only the first
term of this series should be employed further. Using this classification,
one defines the reduced amplitudes $\alpha_1$, $\alpha_2$, $\alpha_3$,
$\alpha_4$ as well as the $B$-functions in the middle point of physical
region of Dalitz-plot at the point $s_0$:

\begin{eqnarray}
s_0^{ik}=0.25(m_i+m_k)^2 s_0 \, ,\nonumber
\end{eqnarray}

\begin{equation}
s_{123}=0.25 s_0 \sum\limits_{i,k=1 \atop i\ne k}^{3} (m_i+m_k)^2 -
\sum\limits_{i=1}^{3} m_i^2 \, ,
\quad s_0=\frac{s+2\sum\limits_{i=1}^{4} m_i^2}
{0.25 \sum\limits_{i,k=1 \atop i\ne k}^{4} (m_i+m_k)^2}
\, . \end{equation}

Such a choice of point $s_0$ allows us to replace integral Eqs. (7) -- (10)
by the algebraic equations (18) -- (21) respectively:

\begin{equation}
\alpha_1=\lambda_1+2\alpha_3 JB_1 (2,2,3)+2\alpha_4 JB_2 (2,2,3)
\, ,\end{equation}

\begin{equation}
\alpha_2=\lambda_2+2\alpha_3 JB_3 (1,1,3)+2\alpha_4 JB_4 (1,1,3)
\, , \end{equation}

\begin{equation}
\alpha_3=\lambda_3+2\alpha_1 JC_1 (3,2,2)+2\alpha_2 JC_2 (3,1,1)
+\alpha_4 JA_1 (3)+\alpha_3 JA_2 (3)\, , \end{equation}

\begin{equation}
\alpha_4=\lambda_4+2\alpha_1 JC_3 (3,2,2)+2\alpha_2 JC_4 (3,1,1)
+\alpha_3 JA_3 (3)+\alpha_4 JA_4 (3)\, . \end{equation}

We use the functions $JA_i(l)$, $JB_i(l,p,r)$, $JC_i(l,p,r)$
$(l,p,r=1-3)$, which are determined by the various $s_0^{ik}$ (Eq. 17).
These functions are similar to the functions:

\begin{eqnarray}
JA_4(l)&=&\frac{G_l^2(s_0^{12})B_l^2(s_0^{23})}
{B_l(s_0^{12})} \int\limits_{(m_1+m_2)^2}^{\frac{(m_1+m_2)^2\Lambda}{4}}
\frac{ds'_{12}}{\pi}\frac{\rho_l(s'_{12})}
{s'_{12}-s_{12}} \int\limits_{-1}^{+1} \frac{dz_1}{2}
\frac{1}{1-B_l (s'_{23})} \, ,\\
&&\nonumber\\
JB_1(l,p,r)&=&\frac{G_l^2(s_0^{12})G_p^2(s_0^{34})B_r(s_0^{23})}
{B_l(s_0^{12})B_p(s_0^{34})}
\int\limits_{(m_1+m_2)^2}^{\frac{(m_1+m_2)^2\Lambda}{4}}
\frac{ds'_{12}}{\pi}\frac{\rho_l(s'_{12})}
{s'_{12}-s_{12}}\nonumber\\
&&\nonumber\\
&\times&\int\limits_{(m_3+m_4)^2}^{\frac{(m_3+m_4)^2\Lambda}{4}}
\frac{ds'_{34}}{\pi}\frac{\rho_p(s'_{34})}{s'_{34}-s_{34}}
\int\limits_{-1}^{+1} \frac{dz_3}{2} \int\limits_{-1}^{+1} \frac{dz_4}{2}
\frac{1}{1-B_r (s'_{23})} \, ,\\
&&\nonumber\\
JC_3(l,p,r)&=&\frac{G_l^2(s_0^{12},\tilde \Lambda)B_p(s_0^{23})
B_r(s_0^{14})}{1- B_l(s_0^{12},\tilde \Lambda)}
\frac{1-B_l(s_0^{12})}{B_l(s_0^{12})}\, \, \frac{1}{4\pi}
\int\limits_{(m_1+m_2)^2}^{\frac{(m_1+m_2)^2\tilde \Lambda}{4}}
\frac{ds'_{12}}{\pi}\frac{\rho_l(s'_{12})}{s'_{12}-s_{12}}\nonumber\\
&&\nonumber\\
&\times&\int\limits_{-1}^{+1}\frac{dz_1}{2}
\int\limits_{-1}^{+1} dz \int\limits_{z_2^-}^{z_2^+} dz_2
\frac{1}{\sqrt{1-z^2-z_1^2-z_2^2+2zz_1z_2}}\nonumber\\
&&\nonumber\\
&\times&\frac{1}{[1-B_p(s'_{23})][1-B_r(s'_{14})]} \, ,
\end{eqnarray}

\begin{eqnarray}
\tilde \Lambda(ik)=\left\{ \Lambda(ik), \hskip5.7em  {\rm if} \quad
\Lambda(ik) \le (\sqrt{s_{123}}+m_3)^2 \atop
(\sqrt{s_{123}}+m_3)^2, \hskip2em {\rm if} \quad
\Lambda(ik) > (\sqrt{s_{123}}+m_3)^2 \right.
\end{eqnarray}

The other choices of point $s_0$ do not change essentially the contributions
of $\alpha_1$, $\alpha_2$, $\alpha_3$ and $\alpha_4$, therefore we omit
the indices $s_0^{ik}$. Since the vertex functions depend only slightly
on energy it is possible to treat them as constants in our approximation.

The solutions of the system of equations are considered as:

\begin{equation}
\alpha_i(s)=F_i(s,\lambda_i)/D(s) \, ,\end{equation}

\noindent
where zeros of $D(s)$ determinants define the masses of bound states of
tetraquarks. $F_i(s,\lambda_i)$ determine the contributions of
subamplitudes to the tetraquark amplitude.

\vskip2ex
{\bf III. Calculation results.}
\vskip2ex
Our calculations do not include the new parameters. We use the cutoff
$\Lambda=7.63$ and the gluon coupling constant $g=1.53$, which are
determined by fixing the tetraquark masses for the states with the hidden
bottom [24, 25]. The widths of the tetraquarks with the open bottom are
fitted by the fixing width $\Gamma_{2^{++}}=(39\pm 26) \, MeV$ [26] for
the $S$-wave tetraquark with the hidden charm $X(3940)$. The quark masses
of model $m_{u,d}=385\, MeV$, $m_s=510\, MeV$ and $m_b=4787\, MeV$ coincide
with our paper ones [25]. The masses and the widths of meson-meson states
with the spin-parity $J^{pc}=0^{++}$ are given in Table I. In our paper we
predicted the tetraquark ($\bar b u \bar u u$) with the mass
$M=5914\, MeV$ and the width $\Gamma_{0^{++}}=104\, MeV$. We calculated
the masses of $X_b(6020)$ $M=6017\, MeV$ and the width
$\Gamma_{0^{++}}=69\, MeV$ (channels $B_s^0 \eta$ and $B^+ K^-$). The
tetraquark ($\bar b s u \bar s$) have the mass $M=6122\, MeV$ and the
width $\Gamma_{0^{++}}=48\, MeV$. The tetraquarks with the open bottom
and the spin-parity $J^{pc}=1^{++}$, $2^{++}$ have only the weak decays.

The functions $F_i(s,\lambda_i)$ (Eq. (26)) allow us to obtain the overlap
factors $f$ for the tetraquarks. We calculated the overlap factors $f$
and the phase spaces $\rho$ for the reactions $X\to M_1 M_2$ (Table I).
We considered the formula $\Gamma \sim f^2 \times \rho$ [27], there $\rho$
is the phase space.

In the open bottom sector the scalar tetraquarks have relatively small
width $\sim 50-100 \, MeV$, so in principle, these exotic states could be
observed.

\vskip2.0ex
{\bf Acknowledgments.}
\vskip2.0ex

The work was carried with the support of the Russian Ministry of Education
(grant 2.1.1.68.26).

\newpage

\noindent
Table I. Masses, widths, overlap factors $f$ and phase spaces $\rho$ of
scalar tetraquarks with open bottom.

\vskip2.5ex

\noindent
\begin{tabular}{|cccccc|}
\hline
Tetraquark & (channels) & $f$ & $\rho$ & Mass ($MeV$) & Widths $(MeV)$
\\[5pt]
\hline
$X_b(5910)$ & $B^+ \eta$ & $0.54$ & $0.103$ & $5914$ & $104$ \\[3pt]
\hline
$X_b(6020)$ &
\begin{tabular}{c}
$B^0_s \eta$\\
$B^+ K^-$
\end{tabular}
&
\begin{tabular}{c}
$0.32$\\
$0.23$
\end{tabular}
&
\begin{tabular}{c}
$0.108$\\
$0.171$
\end{tabular}
& $6017$ & $69$ \\[3pt]
\hline
$X_b(6120)$ & $B^0_s K^+$ & $0.283$ & $0.174$ & $6122$ & $48$ \\[3pt]
\hline
\end{tabular}

\newpage
{\bf \Large References.}
\vskip5ex
\noindent
1. S.K. Choi et al. (Belle Collaboration), Phys. Rev. Lett. {\bf 91},
262001 (2003).

\noindent
2. D. Acosta et al. (CDF Collaboration), Phys. Rev. Lett. {\bf 93},
072001 (2004).

\noindent
3. V.M. Abazov et al. (D0 Collaboration), Phys. Rev. Lett. {\bf 93},
162002 (2004).

\noindent
4. B. Aubert et al. (BaBar Collaboration), Phys. Rev. D{\bf 71},
071103 (2005).

\noindent
5. K. Abe et al. (Belle Collaboration), Phys. Rev. Lett. {\bf 98},
082001 (2007).

\noindent
6. S. Godfrey and N. Isgur, Phys. Rev. D{\bf 32}, 189 (1985).

\noindent
7. L. Maiani, F. Piccinini, A.D. Polosa and V. Riequer,
Phys. Rev. D{\bf 71}, 014028

(2005).

\noindent
8. L. Maiani, A.D. Polosa and V. Riequer, Phys. Rev. Lett. {\bf 99},
182003 (2007).

\noindent
9. D. Ebert, R.N. Faustov and V.O. Galkin, Phys. Lett. B{\bf 634}, 214
(2006).

\noindent
10. L. Maiani, A.D. Polosa and V. Riequer, arXiv: 0708.3997 [hep-ph].

\noindent
11. S.M. Gerasyuta and V.I. Kochkin, arXiv: 0809.1758 [hep-ph].

\noindent
12. S.M. Gerasyuta and V.I. Kochkin, Z. Phys. C{\bf 74}, 325 (1997).

\noindent
13. S.M. Gerasyuta and V.I. Kochkin, arXiv: 0804.4567 [hep-ph].

\noindent
14. S.M. Gerasyuta and V.I. Kochkin, arXiv: 0810.2458 [hep-ph].

\noindent
15. O.A. Yakubovsky, Sov. J. Nucl. Phys. {\bf 5}, 1312 (1967).

\noindent
16. S.P. Merkuriev and L.D. Faddeev, Quantum scattering theory for system
of few

particles (Nauka, Moscow 1985) p. 398.

\noindent
17. Y. Nambu and G. Jona-Lasinio, Phys. Rev. {\bf 122}, 365 (1961):
ibid. {\bf 124}, 246

(1961).

\noindent
18. T. Appelqvist and J.D. Bjorken, Phys. Rev. D{\bf 4}, 3726 (1971).

\noindent
19. C.C. Chiang, C.B. Chiu, E.C.G. Sudarshan and X. Tata,
Phys. Rev. D{\bf 25}, 1136

(1982).

\noindent
20. V.V. Anisovich, S.M. Gerasyuta, and A.V. Sarantsev,
Int. J. Mod. Phys. A{\bf 6}, 625

(1991).

\noindent
21. A.De Rujula, H.Georgi and S.L.Glashow, Phys. Rev. D{\bf 12}, 147 (1975).

\noindent
22. S.M. Gerasyuta and V.I. Kochkin, Yad. Fiz. {\bf 59}, 512 (1996)
[Phys. At. Nucl. {\bf 59},

484 (1996)].

\noindent
23. V.V. Anisovich and A.A. Anselm, Usp. Phys. Nauk. {\bf 88}, 287 (1966)
[Sov. Phys.

Usp. 9, 117 (1966)].

\noindent
24. Y. Cui, X.-L. Chen, W.-Z. Deng and S.-L. Zhu, High Energy Phys. Nucl.
Phys.

{\bf 31}, 7 (2007).

\noindent
25. S.M. Gerasyuta and V.I. Kochkin, arXiv: 0812.0315 [hep-ph].

\noindent
26. C. Amsler et al. (Particle Data Group),
Phys. Lett. B{\bf 667}, 1 (2008).

\noindent
27. J.J. Dudek and F.E. Close, Phys. Lett. B{\bf583}, 278, (2004).

\end{document}